\documentstyle[12pt,epsfig]{article}
\newcommand{\blankline}{\vskip .3cm}
\newcommand{\beq}{\begin{equation}}
\newcommand{\eeq}{\end{equation}}
\newcommand{\tr}{{\mbox tr}}
\renewcommand{\thefootnote}{\fnsymbol{footnote}}
\makeatletter
\makeatletter
\newcommand{\ps@preprint}
\makeatother
\begin{document}
\centerline{\LARGE Planck-scale models of the Universe} 
\blankline
\blankline \centerline{
Fotini Markopoulou\footnote{fotini@perimeterinstitute.ca} }
\blankline
\centerline{\it Perimeter Institute for Theoretical Physics}
\centerline{\it 35 King Street North, Waterloo, Ontario N2J 2W9,
Canada} 
\centerline{\it and} 
\centerline {\it Department of
Physics, University of Waterloo} 
\centerline{\it Waterloo, Ontario
N2L 3G1, Canada} 
\blankline \blankline 
\centerline{October 22, 2002}

\addtocounter{footnote}{-1}
\renewcommand{\thefootnote}{\arabic{footnote}}
%%%%%%%%%%%%%%%%%%%%%%%%%%%%%%%%%%%%%%%%%%%%%%%%%%%%%%%%%%%%%%%%

\section{Introduction}

Suppose the usual description of spacetime as a 3+1 manifold breaks
down at the scale where a quantum theory of gravity is expected to
describe the world, generally agreed to be the Planck scale,
$l_{p}=10^{-35}m$.  Can we still construct sensible theoretical models
of the universe?  Are they testable?  Do they lead to a consistent
quantum cosmology?  Is this cosmology different than the standard one? 
The answer is yes, to all these questions, assuming that quantum
theory is still valid.  After eighty years work on quantum gravity, we
do have the first detailed models for the microscopic
structure of spacetime: spin foams.

The first spin foam models\footnote{Reisenberger, 1994, 1997;
Reisenberger and Rovelli, 1997; Markopoulou and Smolin, 1997; Baez,
1998.} were based on the predictions of loop quantum gravity, namely
the quantization of general relativity, for the quantum geometry at
Planck scale.  A main result of loop quantum gravity is that the
quantum operators for spatial areas and volumes have discrete
spectra\footnote{Rovelli and Smolin, 1995.  For a recent detailed
review of loop quantum gravity see Thiemann, 2001, and for a
non-technical review of the field see Smolin, 2001.}.  Discreteness is central
to spin foams, which are {\em discrete models} of spacetime at Planck
scale.  Several more models have been proposed since, based on results from
other approaches to quantum gravity, such as Lorentzian path
integrals\footnote{ Ambj{\o}rn and Loll, 1998; Loll, 2001; Ambj{\o}rn,
Jurkiewicz and Loll, 2002; Barrett and Crane, 2000; Perez and Rovelli,
2001a.}, euclidean general relativity\footnote{
Barrett and Crane, 1998; Iwasaki, 1999; Perez and Rovelli, 2001b.},
string networks\footnote{ Markopoulou and Smolin, 1998b.}, or
topological quantum field theory\footnote{ See Baez,
2000.}.  For reviews of spin foams see Baez, 2000; Oriti,
2001.

Spin foam models are background independent, i.e.\ they do not live in a
pre-existing spacetime.  Gravity and the
familiar 3+1 manifold spacetime are to be derived as the low-energy
continuum approximation of these models.  Thus, a spin foam model will
be a good candidate for a quantum theory of gravity only if it can be
shown to have a good low energy limit which contains the known
theories, namely, general relativity and quantum field theory.  One
also expects a good model to predict observable departures from these
theories.

Spin foams are only a few years old, and progress towards finding
their low-energy limit is still in its very early stages.  The aim of
this note is both to discuss the basic features of these models, as
well as their current status and ways to proceed in future research. 
Section 2 contains mostly results on the general formalism of these
models.  We see that they lead to a novel description of the universe,
including a consistent quantum cosmology, in which, in general, there
is no wavefunction of the universe or Wheeler-DeWitt equation.  In
section 3, we note that every spin foam is given by a partition
function very similar to that of a spin system or a lattice gauge
theory.  I argue that this suggests that we should treat this approach
to quantum gravity as a problem in statistical physics.  However,
there is an important difference from systems in statistical physics,
the background-independence of spin foams.  I list features of spin
foams relevant to the calculation of their low-energy limit and
discuss ways to proceed.  Finally, spin foams can address the current
challenge that quantum gravity effects, such as breaking of Lorentz
invariance, may be observable\footnote{For  possible
experiments probing quantum gravity effects see Jacobson, Liberati and
Mattingly, 2001; Sarkar, 2002; Amelino-Camelia, 2002; Konopka and
Major, 2002; Kempf, 2002; Ellis et al.\, 2002.}.  
In Section 3 and the Conclusions, we discuss the kind
of predictions one could calculate with these models.

%%%%%%%%%%%%%%%%%%%%%%%%%%%%%%%%%%%%%%%%%%%%%%%%%%%%%%%%%%%%%%%%

\section{No spacetime manifold $+$ Quantum theory $=$ Spin foams}

Several models of the microscopic structure of spacetime have recently 
been proposed.  Different ones were constructed based on 
different motivations, but they have several features in common which 
I list here\footnote{This is a 
rather personal interpretation of spin foam models.  Several of the 
models in the spin foam literature are constructed as a path integral 
formulation of loop quantum gravity, or are modelled on topological 
quantum field theory, and are not causal, nor is discreteness always 
considered fundamental.  For reviews of spin foams from alternative 
viewpoints, see Baez, 2000; Oriti, 2001.}:
\begin{description}
    \item[A]  
    At energies close to the Planck scale, the description of
    spacetime as a 3+1 continuum manifold breaks down.  This is the
    old explanation for the singularities of general relativity and is
    further supported by the results of loop quantum gravity.

    \item[B]  At such energies the universe is discrete.  This is a simple way 
    to model the idea that in a finite region of the universe there 
    should be only a finite number of fundamental degrees of freedom.  
    This is supported by Bekenstein's arguments, by
    the black hole entropy calculations from both 
    string theory and loop quantum gravity, by the quantum geometry 
    spectra of loop quantum gravity, and is related to holographic 
    ideas.

    \item[C]  
    Causality still persists even when there is no manifold spacetime. 
    How to describe a discrete causal
    universe has been known for quite some time, it is 
    a {\em causal set} (Bombelli et al., 1987; Sorkin, 1990).  This is a set of
    events $p, q, r, \ldots$ ordered by the causal relation $p\leq q$, 
    meaning ``$p$ preceeds $q$'',
    which is transitive ($p\leq q$ and $q\leq r$ implies that $p\leq
    r$), locally finite (for any $p$ and $q$ such that $p\leq q$, the
    intersection of the past of $q$ and the future of $p$ contains a
    finite number of events), and has no closed timelike loops (if
    $p\leq q$ and $q\leq p$, then $p=q$).  Two events $p$ and $q$ are
    unrelated (or spacelike) if neither $p\leq q$ nor $q\leq p$ holds.
    
    Note that the microscopic events do not need to be the same (or a 
    discretization of) the events in the effective continuum theory.  
    Also, the speed of propagation of information in the microscopic 
    theory does not have to be the effective one, the speed of light 
    $c$.

    \item[D] Quantum theory is still valid. 
    
    \item[E]  Since we are modelling the universe itself, the model should be 
    background independent.
\end{description}

\begin{figure}
\center{
\epsfig{file=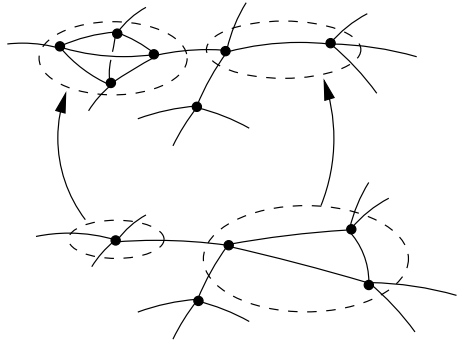}}
\caption{One evolution step in a causal spin network model.  Only parts of 
the spin networks are shown.  }
\label{csn}
\end{figure}

An example of such a model is {\em causal spin networks} (Markopoulou
and Smolin, 1997; Markopoulou, 1997).  Spin networks were originally
defined by Penrose as trivalent graphs with edges labelled by
representations of $SU(2)$ (Penrose, 1971).  From such abstract labelled
graphs, Penrose was able to recover directions (angles) in
3-dimensional Euclidean space in the large spin limit.  Later, in loop
quantum gravity, spin networks were shown to be the basis states for
the spatial geometry states.  The quantum area and volume operators,
in the spin network basis, have discrete spectra, and their
eigenvalues are functions of the labels on the spin network.

Given an initial spin network, to be thought of as modeling a quantum
``spatial slice'', a causal set is built by repeated application of
local moves, local changes of the spin network graph.  Each move 
results in a causal relation in the causal set.  
An example is shown in Fig.\ref{csn}.

One can show that a small set of local generating moves can be 
identified that take us from any given network to any other one. 
For example, for 4-valent networks, we only need the 
following local moves on pieces of the network: 

\[
    \begin{array}{c}\mbox{\epsfig{file=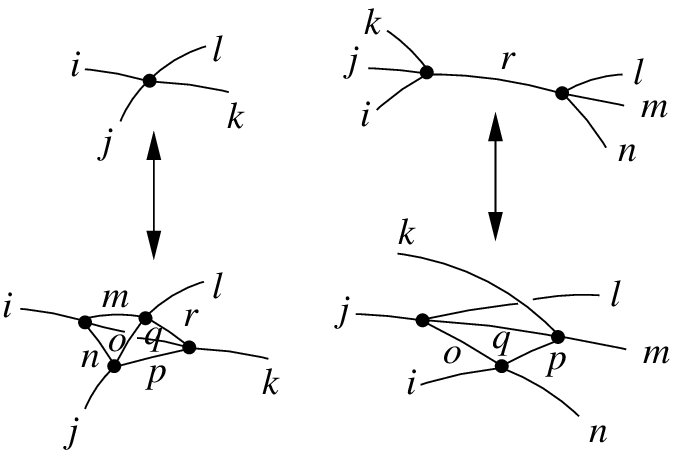}}\end{array}
\]

One should note that there is no preferred foliation in this model. 
The allowed moves change the network locally and any foliation
consistent with the causal set (i.e.\ that respects the order the
moves occured) is possible.  This is a discrete analogue of
multifingered time evolution.  For more details, see Markopoulou,
1997.

There is an amplitude $A_{\mbox{\footnotesize{move}}}$ for each move 
to occur.  The 
amplitude to go from a given initial spin network $S_{i}$ to a final 
one $S_{f}$
is the product of the amplitudes for the generating moves in the 
interpolating history of sequence of such moves, summed over all possible 
such sequences (or histories):
\beq
A_{S_{i}\rightarrow S_{f}}=\sum_{{\mbox{\footnotesize histories }}S_{i}\rightarrow 
S_{f}}
\prod_{\stackrel{\mbox{\footnotesize moves}}
{\mbox{\footnotesize in history}}} A_{\mbox{\footnotesize move}}.
\label{Zcsn}
\eeq 
Explicit expressions for the amplitudes $ A_{\mbox{move}}$ have
so far been given in Borissov and Gupta, 1998, for a simple causal
model, in Ambj{\o}rn and Loll, 1998 (and their higher dimensional
models), with differences in the allowed 2-complexes, and very recently
in Livine and Oriti, 2002 for the Lorentzian Barrett-Crane
model.

\subsection{The general formalism of the models}

With this example in mind, we can write down the formalism of the 
generic model that has the features {\bf A}-{\bf E} above.  This will 
let us derive results about the 
general form of $A_{\mbox{\footnotesize move}}$ and the resulting 
quantum cosmology. 

In the particular example of 
 the causal spin networks, we note that the model really is a
causal set ``dressed with quantum theory'' as follows: A move in the
history changes a subgraph of the spin network with free edges.  To
such a subgraph $s$ is naturally associated a Hilbert space $H$ of 
so-called intertwiners.  These are maps from the tensor product of the 
representations of $SU(2)$ on the free edges to the identity representation.  

\begin{figure}
\center{
\epsfig{file=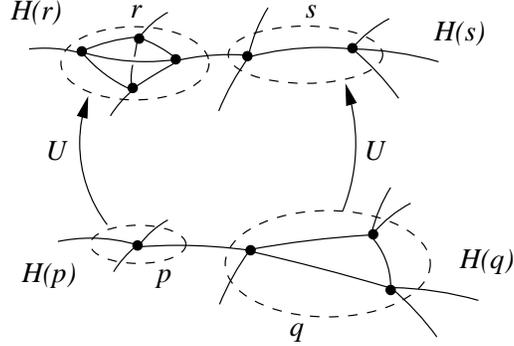}}
\caption{A subgraph in the spin network corresponds to a Hilbert 
space of intertwiners.  It is also an event in the causal set.  Two 
spacelike events are two independent subgraphs, and the joint Hilbert 
space is $H(p\cup q)=H(p)\otimes H(q)$ if they have no common edges, 
or $H(p\cup q)=\sum_{i_{1},\ldots i_{n}}H(p)\otimes H(q)$, if they are joined by 
$n$ edges carrying representations $ i_{1},\ldots i_{n}$.  In this 
example,  $H(p\cup q)=\sum_{m}H(p)\otimes H(q)$. }
\label{Qcsn}
\end{figure}

The new subgraph, $s'$ has the same boundary as $s$, the same edges 
and labels, and therefore corresponds to the same Hibert space of 
intertwiners.  A move is a unitary operator from a state 
$|\Psi_{s}\rangle$ to a new one $|\Psi_{s'}\rangle$ in $H$.  See 
Fig.\ref{Qcsn}.  

Therefore, a causal spin network history is a causal set in which the 
events are Hilbert spaces and the causal relations are unitary 
operators. 

Is it true for any model with properties {\bf A}-{\bf E} that it 
is such an assignment of a quantum theory to a causal set?  
The answer is yes, although the assignment can be slightly more 
complicated than for causal spin networks.  

We start by interpreting events in $C$ as the smallest Planck scale
systems in the quantum spacetime.  These, according to {\bf D}, are
quantum mechanical.  Quantum theory describes the possible states of
such a system as states in a Hilbert space, if it is an isolated
system, or by a density matrix in the more general case of an open
system.  It turns out that in our models each event $p\in C$ is best
described by a density matrix $\rho(p)$ (for reasons we will explain
below).

\begin{figure}
\center{
\epsfig{file=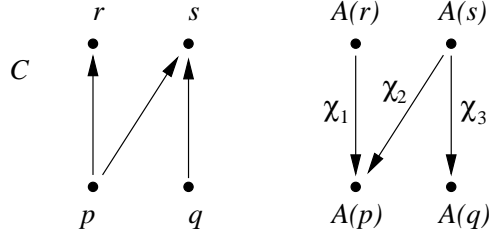}}
\caption{On the left we have four events $p,q,r,s$ in a causal set and 
their causal relations.  On the right are the  
corresponding density matrices and quantum operations $\chi$ 
in the quantum version.  }
\label{ach}
\end{figure}

Going from a causal set $C$ to a quantum spacetime then involves the following
steps: a) To each event $p\in C$ we assign an algebra of operators
$A(p)$.  Property {\bf B} implies that $A(p)$ is a simple matrix
algebra.  Any such algebra carries a unique, faithful, normal trace
$\tau:A\rightarrow {\bf C}$ defined by the properties that
$\tau(ab)=\tau(ba)$ and $\tau(1)=1$, and given by the formula
$\tau(a)=\tr a/\tr 1$.  This makes the algebra into a
finite-dimensional Hilbert space with inner product $\langle a|\rangle
b:=\tau(a^{b})$. 
b) The density matrix $\rho(p)$ representing the
state of $p$ is a positive-definite operator in $A(p)$.  c) Two
spacelike events $p$ and $q$ are two independent events, and so are in
a composite state given by $\rho(p)\otimes \rho(q)$.  
d) Every causal
relation $p\leq q$ in the causal set corresponds to a {\em quantum
operation} $\chi:A(q)\rightarrow A(p)$ (Fig.\ref{ach}).  This is the most
general physical transformation that quantum theory allows between two
open systems.  A quantum operation is a completely positive linear
operator, namely: it is linear on the $\rho$'s, it is trace-preserving
($\tr(\rho)=\tr(\rho')=1$), positive, and completely positive (if 
$\chi:\rho(q)\rightarrow \rho(p)$ is positive, then $\chi\otimes {\bf 
1}: \rho(q)\otimes \rho(s)\rightarrow \rho(p)\otimes \rho(s) $ is also 
positive).

\begin{figure}
\center{
\epsfig{file=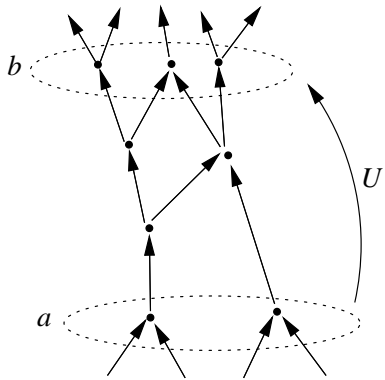}}
\caption{The acausal sets $a$ and $b$ form a complete pair.}
\label{complete}
\end{figure}

We now have a formalism of models with properties {\bf A}-{\bf E} as 
a collection of density matrices connected by quantum operations.  
When can we have unitary evolution in this quantum spacetime?
To answer 
this, let us first define an {\em acausal set}.  This is a subset 
$a=\{p,q,r,\ldots\}$ of $C$ with $p,q,r,\ldots$ all spacelike to each 
other.  It is not difficult to check that  {\em unitary evolution is only 
possible between two acausal sets $a$ and $b$ that form a 
complete pair}, namely, every event in $b$ is in the future of some 
event in $a$ and every event in $a$ is in the past of some event in 
$b$.  This is because, by construction, information is conserved from $a$ to $b$. See the 
example in
Fig.\ref{complete}.

The fundamental description of the quantum spacetime as a
collection of open systems joined by quantum operations does
contain all the relevant physics, including the causal relations and
any unitary operations.  It is a rather technical construction to
discuss here, but one can show that the causal information of $C$ is
contained in conditions on the quantum operations, and can prove that
given a complete pair $a$ and $b$, the quantum operations on the
causal relations interpolating from $a$ to $b$ compose to
give precisely the unitary transformation $U:A(a)\rightarrow A(b)$
(Hawkins, Markopoulou and Sahlmann, 2002).

Therefore, our quantum spacetime is a very large set of open systems
connected by quantum operations, where unitary evolution arises only
as a special case, for a complete pair (the special case of an
isolated system).  It is interesting to note, first, that this
description of a quantum spacetime is almost identical to the quantum
information theoretic description of noise in a quantum operation
(e.g.\ see Nielsen and Chuang, 2000, p.353), and, second, that a
master equation, already extensively used in quantum cosmology, is a
continuum an analogue of a quantum operation (as above, p.386.).

 These quantum spacetimes lead us to a new quantum cosmology which we
 describe next.

\subsection{Quantum cosmology}

The standard quantum cosmology is based on the recipe for the 
canonical, or 3+1, quantization of gravity.  Here one starts with a 
spacetime with the topology $\Sigma\times R$, where $\Sigma$ is the 
3-dimensional spatial manifold.  Quantizing the geometry of $\Sigma$ 
(identifying variables such as the 3-metric and extrinsic curvature, 
or Ashtekar's new variables) we obtain the so-called {\em 
wavefunction of the universe} $|\Psi_{\mbox{\footnotesize univ}}\rangle$.  
An example of such a state is the Chern-Simons state in loop quantum 
gravity. This is 
to ``evolve'' eccording to the Wheeler-DeWitt equation, 
\beq
\widehat{H}|\Psi_{\mbox{\footnotesize univ}}\rangle=0,
\label{WDW}
\eeq
where $\widehat{H}$ is the quantization of the Hamiltonian constraint 
in the 3+1 decomposition of the Hilbert-Einstein action of general 
relativity, a hermitian operator.  

\begin{figure}
\center{
\epsfig{file=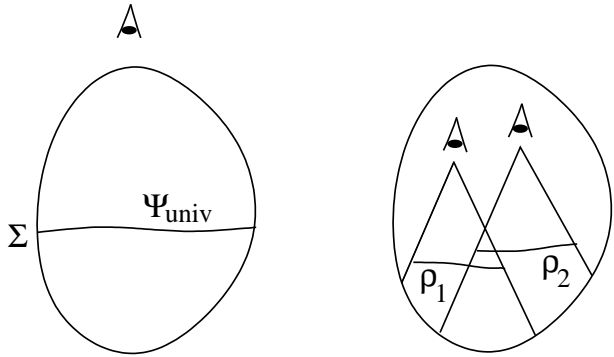}}
\caption{A wavefunction of the universe can only be seen by an observer outside 
the universe.  A quantum theory of cosmology should refer to 
observables measurable from inside the universe.  Inside observers 
have only partial information of the universe, since only events in 
their causal past are accessible to them.}
\label{psi}
\end{figure}

There are several issues with this.  First, the simple form of the
equation and especially the peculiar righthand side hides the fact
that we need to quantize relativity, a background-independent theory. 
We only really understand the quantum mechanical evolution of ordinary
systems, where an external time is always unambiguously present. 
Second, equation
(\ref{WDW}) only works for spacetimes that are globally hyperbolic. 
Third, one can argue that $|\Psi_{\mbox{\footnotesize univ}}\rangle$
and eq.\ (\ref{WDW}) do not have a satisfactory physical
interpretation:  $|\Psi_{\mbox{\footnotesize univ}}\rangle$ is the
state of the entire universe and thus only accessible to an observer
{\em outside} the universe (or specific observers in special
universes, such as the final moment of a spacetime with a final
crunch, etc).  A satisfactory theory of quantum cosmology has,
instead, to refer to physical observations that can be made from
inside the universe  (Markopoulou, 1998) (see the diagram in Fig.\ref{psi}).

In the miscroscopic models we defined, the analogue of a spatial 
slice is an acausal set that is maximal, namely a subset of $C$ such 
that every other event in $C$ is either in its past or in its 
future.  By tensoring together all the density matrices on each event 
in this ``slice'', we could obtain a microscopic 
$|\Psi_{\mbox{\footnotesize univ}}\rangle$.  However, the causal 
structure of the generic $C$ is very different than that of a 
globally hyperbolic spacetime.  One can show that, on average, a 
generic $C$ has very few ``slices'' (Meyer, 1988).  And these may cross, 
i.e.\ one is partly to the future and partly to the past of the 
other.  All this makes $|\Psi_{\mbox{\footnotesize univ}}\rangle$ and 
the WDW equation not very useful for the generic causal 
set.  We cannot restrict to causal sets that admit foliations since 
these are very special configurations in the partition function of 
the models. 

\begin{figure}
\center{
\epsfig{file=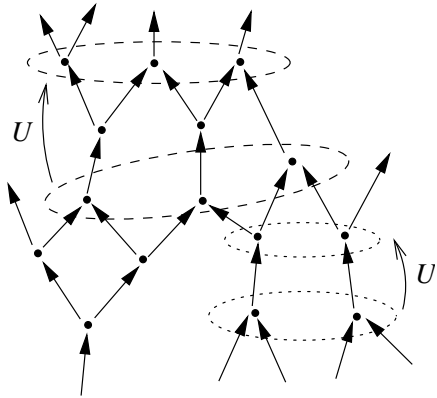}}
\caption{Two complete pairs in a quantum spacetime. }
\label{cpairs}
\end{figure}

The interesting fact is that the models provide an alternative quantum
cosmology that does not use a wavefunction of the universe and in fact
avoids the issues listed above.  The universe is not represented by a
global $|\Psi_{\mbox{\footnotesize univ}}\rangle$, but is instead a
collection of ordinary open quantum mechanical systems (all the
density matrices on the events of $C$).  Or, at the level of complete
pairs, it is a collection of ordinary isolated quantum mechanical
systems.  There is no WDW equation, but there is {\em local unitary
evolution} and a partition function for the entire system (see
Fig.\ref{cpairs}). 
These local systems may or may not combine to give
an evolving wavefunction of the universe, depending on the causal
structure.  As a result, 
any observables naturally refer to observations made from
inside the universe (see Hawkins, Markopoulou and Sahlmann, 2002).

A smooth continuous universe with $\Sigma\times R$ topology is what we 
want to derive in the low-energy limit.  Viewed this way, 
$|\Psi_{\mbox{\footnotesize univ}}\rangle$ and the WDW equation 
presupposes the limit we need to derive. 

%%%%%%%%%%%%%%%%%%%%%%%%%%%%%%%%%%%%%%%%%%%%%%%%%%%%%%%%%%%%%%%%

\section{Quantum Gravity as a problem in statistical 
physics}

We now wish to discuss the problem of calculating the low-energy limit 
of spin foam models.  To do so, we give the general definition of a 
spin foam, a partition function of which equation (\ref{Zcsn}) is a 
special case. 

\begin{figure}
\center{
\epsfig{file=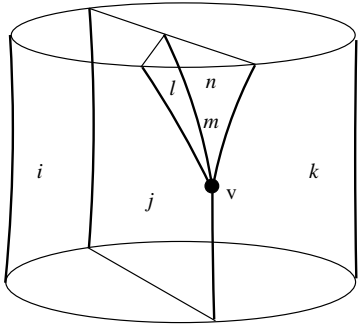}}
\caption{A spin foam is a 2-complex whose faces are labelled by group 
representations as shown.  Cuts through a spin foam are spin networks 
(graphs with edges labelled by the representations on the faces they 
cut through). Vertices, such as $v$ above, correspond to the moves in 
the causal spin network example.  }
\label{sfoam}
\end{figure}

A spin foam is a labelled 2-complex whose faces carry
representations of some group $G$, the edges by intertwiners in the 
group, and the vertices carry the evolution amplitudes.  These are 
functions of the faces and the edges that meet on that vertex and 
code the evolution dynamics for the model (fig.\ref{sfoam}).  

A spin foam model is then given by a partition function of the form
\beq
Z(S_{i},S_{f})=\sum_{\Gamma}
           \sum_{\mbox{labels on }\Gamma}\prod_{f\in\Gamma}
	   \mbox{dim}j_{f}\prod_{v\in\Gamma}A_{v}(j).
	   \label{Z}
\eeq
The first sum is over all spin foams $\Gamma$ interpolating between a 
given initial spin network $S_{i}$ and a final one $S_{f}$.  
$\mbox{dim}j_{f}$ is the dimension of the $G$ representation labelling 
the face $f$ of $\Gamma$.  $A_{v}$ is the amplitude on the vertex of 
$v$ of $\Gamma$, a given function of the labels on the faces and the 
edges adjacent to $v$.  A choice of the group $G$ and the functions 
$A_{v}$ (and possibly a restriction on the allowed 2-complexes) 
defines a particular spin foam model.   Degrees of 
freedom such as matter and supersymmetric ones can be introduced by 
using or adding the appropriate group representations.  

Several spin foam models exist in the literature, all candidates for
the microscopic structure of spacetime.  The very first test such a
model has to pass is a tough one: it should have a good low energy
limit, in which it reproduces the known theories, general relativity
and quantum field theory.  Next, it should predict testable deviations
from these theories.  The first question is then, what is this limit
and how we are going to calculate it.

Note that the models are given by a partition function that is
strongly reminiscent of that for a spin system or a lattice gauge
theory.  This suggests that the problem of their low-energy limit may
be best treated as a problem in statistical physics.  That is, for
spin foams in the correct class of microscopic models, we should find
the known macroscopic theory by integrating out microscopic degrees of
freedom in $Z$.

Can we use techniques from statistical physics to test the models? 
This appears very promising, however, there are important differences
between spin foams and the systems studied in condensed matter
physics. 
We can actually list the features of spin foam models, so that we can 
make a better comparison with the situation in condensed matter 
physics.  They are:
\begin{enumerate}
    \item 
    The microscopic degrees of freedom are representations of a group 
    or algebra.  
    \item
    The weights in the partition 
    function are amplitudes rather than probabilities. 
    \item
    The lattices are the highly irregular spin foam 2-complexes.  In 
    general, we cannot simplify the problem by restricting to regular 
    2-complexes as these are rare configurations in the sum. 
    \item
    Spin foams are background-independent.  This means that we cannot 
    use global external parameters such as time or temperature.  
    \item
    There is a minimum length, the Planck length. 
    \item 
    The partition function contains a {\em sum} over all 2-complexes 
    with the same given boundary.
\end{enumerate}
1, 2 or 3 above are mainly technical difficulties.  4, however, and 6, 
are novel issues, due to the fact that spin foams are microscopic 
models of the universe itself.   It is possible that one 
can extend the methods of statistical physics, such as the 
renormalization group, to deal with background independent systems 
(Markopoulou, 2002).  

One thing that is true in statistical physics is that progress is made
by analysing specific models, and the issues mentioned above may or
may not turn out to be significant.  For example, a very interesting
model is the Ambj{\o}rn-Loll model of Lorentzian dynamical
triangulations (Ambj{\o}rn and Loll, 1998).  In this model, quantum 
spacetimes of piecewise linear simplicial building blocks approximate 
continuum Lorentzian spacetimes.  To reflect  the causal properties 
of the continuum spacetimes, the model does not allow any spatial 
topology change.  As a result, it has a foliated structure.  It is 
easy to describe this in 1+1 dimensions.  There, the model is a sum 
over sequences of 
discrete one-dimensional spatial slices, namely, closed chains of 
length $L$ that changes in time.

From the perspective of a relativist, for whom explicit background
independence is a necessary condition for a model of the universe to
be satisfactory, this model is unpleasant because it appears to have a
preferred foliation.  The relativist will also question the exclusion
of topology change.  However, the model is completely well-defined,
and the suppression of topology change enables us to perform a Wick
rotation, solve it analytically and find that it has a good low-energy
limit with very interesting properties.  This limit cannot, of course,
be classical gravity, since general relativity in 1+1 is an empty
theory.  Still, one finds that this model belongs to a different universality
class than the well-studied euclidean (Liouville) 2-d quantum gravity,
and that it is much better behaved.  For example, its Hausdorff
dimension\footnote{The Hausdorff dimension $d_{H}$ can be measured by
finding the scaling behaviour of the volumes $V(R)$ of geodesic balls
of radius $R$ in the ensemble of Lorentzian geometries: $\langle
V(R)\rangle\sim R^{d_{H}}$.} is 2, compared to the result for
euclidean 2-d histories, which have Hausdorff dimension 4 (reflecting 
the dominance of fractal geometries).  The
physically reasonable result of 2 is a direct consequence of the
suppression of topology change and the resulting foliated structure.

There are similar results for these models in higher dimensions (see
Loll, 2001).  Certainly, we cannot have a final verdict on this model
until we have its solution in four dimensions.  However, it raises the
possibility that something already familiar from statistical physics,
namely, that the properties of the low-energy theory do not have to be
present in the microscopic model, may hold even for
backround-independence.

If we regard a spin foam as a statistical physics model, then the
phenomenon of universality suggests that it is very likely that models
with different microscopic details have the same low-energy limit. 
This is in contrast with many current arguments for or against
specific spin foam models.  Most spin foam models are derived from
other approaches to quantum gravity (such as path-intergal form of
loop quantum gravity, deformations of topological quantum field
theories, etc.), and so there is attachment to the details of the
models.  For example, a very popular model is the Lorentzian
Barrett-Crane model (Barrett and Crane, 2000).  It has a partition
function of the form (\ref{Z}), with representations of the Lorentz
group.  We know that the Lorentz group is present in the observed
low-energy theory (as opposed to Euclidean gravity which is a
mathematical construct).  This is taken to mean that this model is
preferred over the Euclidean Barrett-Crane model (Barrett and Crane,
1998).  But what is the status of this choice if the Lorentz group
appears only in the low-energy theory?

It is my personal opinion that such arguments, for or against
particular details of the partition function (\ref{Z}), at this stage,
miss the point.  What is now required is calculations of collective
effects in a spin foam. For example, what many spin foam models
suppose is that there exist discrete fundamental building blocks of
spacetime.  This is more striking than the details of these
blocks.  Can we demonstrate their existence independently of their 
detailed structure?  

This brings us to the second lesson from statistical physics:
experiments are necessary.  We currently have several proposals for
experiments that will probe the high-energy regime of spacetime.  I
believe the task at hand is to make contact between the partition
function (\ref{Z}) and such experiments.  Calculation of collective
effects in a spin foam can be used to predict, for example, departures
from Lorentz incariance.  This is not an easy task considering the
great gap from the Planck scale to what is currently accessible
experimentally, and it is further complicated by questions about what
time, temperature, etc.\ are in these models.  But the upside is that,
if this works, we will have testable real-physics quantum gravity.

%%%%%%%%%%%%%%%%%%%%%%%%%%%%%%%%%%%%%%%%%%%%%%%%%%%%%%%%%%%%%%%%

\section{Conclusions}

In the last few years, spin foam models have been proposed as the 
microscopic description of spacetime at Planck scale.  We have 
described their general features and we gave the general formalism for 
such models.  

Causal spin foam models provide a new quantum cosmology in which 
there is no wavefunction of the universe or Wheeler-DeWitt equation.  
Instead the universe is described simply as a collection of ordinary 
quantum mechanical systems.  

Spin foams are best interpreted as {\em models}\/ of the universe in 
the statistical physics sense, with gravity and 3+1 spacetime to be 
derived as the low-energy continuum limit (although this is not the 
way most were introduced).  I believe our best chance to calculate 
their continuum limit is indeed by importing methods from statistical 
physics.  This immediately implies two things: a) the microscopic 
details of the models may not play a role, and b) further progress 
should be made by analyzing individual models as well as by 
experimental input.  

It is tempting to compare our current situation to the 1900's, shortly
before atomism was established.  It is hard to believe it now, but at
the time, the idea that there could be ``{\em any} hypothesis about
the microstructure of matter was opposed on the grounds that (a) such
a structure is inherently unobservable; (b) phenomenological theories
are quite adequate for the legitimate purposes of science'' (quoted 
from Brush, 
1986, p.\ 92).  Many
models of the atomic theory of matter were proposed at the time. 
``Every Tom, Dick and Harry felt himself called upon to devise his own
special combination of atoms and vortices, and fancied in having done
so that he had pried out the ultimate secrets of the
Creator''\footnote{L.\ Boltzmann, {\em Verh.\ Ges.\ D.\ Naturf.\ 
Aerzte}\/ (1) 99 (1899).  Quoted from Brush, 1986, p.\ 98.}.  It is interesting to see how
misguided the physicists were as to the abilities of the
experimentalists.  They did not think that atoms would be observable
in their lifetime.  It is also interesting to see that the proof that
atoms exist did not involve any particular model.  It came with
Einstein's theory of Brownian motion, where very basic assumptions on
the statistical nature of molecules (that they are identical,
interchangeable, etc), allowed him to calculate a collective effect,
that the mean disctance travelled by a molecule was proportional to
the square root of time, which was {\em observable} and {\em
different} from the corresponding result for continuous matter.

Again, experience from statistical physics teaches us that
experimental input is required to identify the correct models for the
systems we are interested in.  We are certainly at an early stage but
we may well be entering a very exciting period for quantum gravity. 
There is a real chance that experimental data will come in in the next
few years that we will have to explain, with spin foam models, or some
other approach.  We can start to treat quantum gravity as real
physics, where we make contact with experiment and compare predictions
with experimental data.

%%%%%%%%%%%%%%%%%%%%%%%%%%%%%%%%%%%%%%%%%%%%%%%%%%%%%%%%%%%%%%%%

\section*{References}
\small{

\begin{verse}
    
Ambj{\o}rn, J.,  and R.\ Loll, 1998,
``Non-perturbative Lorentzian Quantum Gravity, Causality and
Topology Change'',
Nucl.Phys.\  B536 407, preprint available as hep-th/9805108;

Ambj{\o}rn, J., J. Jurkiewicz and R. Loll, ``3d Lorentzian, Dynamically Triangulated 
Quantum Gravity'', Nucl.Phys.Proc.Suppl.\ 106 (2002) 980-982, preprint 
available as hep-lat/0201013. 

Amelino-Camelia, G., ``Quantum-Gravity Phenomenology: Status and Prospects''
Mod.Phys.Lett. A17 (2002) 899-922, preprint available as gr-qc/0204051.

Baez, J.\, ``Spin Foam Models'', Class.\ Quant.Grav.\ 15 (1998) 
1827-1858, preprint available as gr-qc/9709052.

Baez, J.\, ``An Introduction to Spin Foam Models of Quantum 
Gravity and BF Theory'', Lect.Notes Phys. 543 (2000) 25-94, preprint 
available as gr-qc/9905087.

Barrett, J., and  L.\ Crane,
``Relativistic spin networks and quantum gravity'',
J.Math.Phys. 39 (1998) 3296-3302, preprint available as 
gr-qc/9709028.

Barrett, J., and  L.\ Crane,
``A Lorentzian Signature Model for Quantum General Relativity'',
Class.Quant.Grav. 17 (2000) 3101-3118, preprint available as gr-qc/9904025.

Bombelli L, Lee J, Meyer D and 
 R, 1987,
{``Space-time as a causal set''}, {\it Phys Rev Lett}\/ {\bf 59}
521.

Borissov R and Gupta S, 1998,
``Propagating spin modes in canonical quantum gravity'', {\it Phys 
Rev} {\bf D 60} (1999) 024002 (gr-qc/9810024).

Brush, S.G., {\em The kind of motion we 
        call heat}\/ Vol.\ 1, 1986 (Amsterdam: North Holland)
	
Ellis, J., N.E.\ Mavromatos, D.V.\ Nanopoulos and A.S.\ Sakharov, 
``Quantum-gravity analysis of gamma-ray bursts using wavelets'', 
astro-ph/0210124.
	
Hawkins, F.\ Markopoulou and H.\ Sahlmann, 
        Algebraic causal histories, to appear. 
	
Iwasaki, J., ``A surface theoretic model of quantum 
gravity'', preprint available as gr-qc/9903112.

Jacobson, T., S.\ Liberati and D.\  Mattingly, 
    ``TeV Astrophysics Constraints on Planck Scale Lorentz Violation'',
    hep-ph/0112207.

Kempf, A., ``On the vacuum energy in expanding space-times'', 
gr-qc/0210077.
    
Konopka, T.J. and  S. A. Major, ``Observational Limits on Quantum Geometry 
    Effects'',  New J.Phys. 4 (2002) 57, hep-ph/0201184.
    
Livine, E.R., and Oriti, D, ``Implementing causality in the spin foam 
quantum geometry'', preprint available as gr-qc/0210064.

Loll, R., ``Discrete Lorentzian Quantum Gravity'', Nucl.\ Phys.\ 
Proc.\ Suppl.\ 94 
(2001) 96-107, preprint available as hep-th/0011194;

Markopoulou, F., ``Dual formulation of spin network evolution'',
preprint available as gr-qc/9704013.

Markopoulou F, 1998, {``The internal logic of causal sets: What the
universe looks like from the inside''},Commun.Math.Phys.  211 (2000)
559-583 (gr-qc/9811053).

Markopoulou,  F., ``Coarse-graining spin foam 
        models'', gr-qc/0203036
 
Markopoulou, F.,  and  L.\ Smolin, ``Causal evolution of spin networks'',
Nucl.Phys.\ B508 (1997) 409, preprint available as gr-qc/9702025.

Markopoulou, F., and L.\ Smolin, ``Quantum geometry with intrinsic
local causality'',
Phys.Rev. D58 (1998) 084032, preprint available as gr-qc/9712067.

Markopoulou, F.\ and L.\ Smolin, ``Nonperturbative dynamics for
abstract (p,q) string networks'', Phys.\ Rev.\ D58 (1998) 084033,
preprint available as hep-th/9712148.

Meyer D A, 1988, {\sl The Dimension of Causal Sets}, PhD Thesis,
Massachussets Institute of Technology.

Nielsen, M.A. and I.L.Chuang, {\em Quantum 
    Computation and Quantum Information}, 2000 (Cambridge: Cambridge 
    University Press).
    
Oriti, D., ``Spacetime geometry from algebra: spin foam models for
non-perturbative quantum gravity'', Rept.Prog.Phys.\ 64 (2001) 
1489-1544, preprint available as
gr-qc/0106091.

Penrose, R., ``Theory of quantized directions'', 
unpublished manuscript, and in ``Quantum theory and beyond'', ed.\ T.\ 
Bastin, Cambridge U.\ Press 1971.

Perez, A., and C.\ Rovelli, ``Spin foam model for Lorentzian 
General Relativity'', Phys.Rev. D63 (2001) 041501, preprint available 
as gr-qc/0009021.

Perez, A., C.\ Rovelli,`` A spin foam model without 
bubble divergences'', Nucl.Phys.\ B599 (2001) 255-282, preprint available 
as gr-qc/0006107.

Reisenberger, M.,
``Worldsheet formulations of gauge theories and gravity'',
preprint available as gr-qc/9412035.

Reisenberger, M., ``A lattice worldsheet sum for 4-d
Euclidean general relativity'', preprint available as
gr-qc/9711052.

Reisenberger, M., and C.\ Rovelli,  ` ``Sum over Surfaces'' form
of Loop Quantum Gravity',
Phys.Rev.\ D56 (1997) 3490, preprint available as gr-qc/9612035.

Rovelli, C,  and L.\ Smolin, ``Discreteness of area and volume in
quantum gravity'', Nucl.\ Phys.\  B442 (1995) 593-622; Erratum-ibid.
B456 (1995) 753, preprint available as gr-qc/9411005.

Sarkar, S., ``Possible astrophysical probes of quantum gravity'',
Mod.\ Phys.\ Lett.\  A17 (2002) 1025-1036, gr-qc/0204092.
    
Smolin, {\em Three roads to quantum gravity}.  London: Weidenfeld and 
Nicholson, 2000.

Sorkin R, 1990, {``Space-time and causal sets''} 
in Proc.\ of SILARG VII Conf., Cocoyoc, Mexico.

Thiemann T., ``Introduction to Modern Canonical Quantum General
Relativity'', preprint available as gr-qc/0110034.

\end{verse}
 
}

\end{document}